\documentclass[twocolumn,showpacs,preprintnumbers,amsmath,amssymb,superscriptaddress]{revtex4}

\usepackage{graphicx}
\usepackage{graphicx}
\begin{document}
\title{Chemical potential jump between hole- and electron-doped sides of ambipolar high-$T_{\rm c}$ cuprate}

\author{M. Ikeda}
\affiliation{Department of Complexity Science and Engineering and Department of Physics, University of Tokyo, Hongo 7-3-1, Bunkyo-ku, Tokyo 113-0033, Japan}

\author{M. Takizawa}
\affiliation{Department of Complexity Science and Engineering and Department of Physics, University of Tokyo, Hongo 7-3-1, Bunkyo-ku, Tokyo 113-0033, Japan}

\author{T. Yoshida}
\affiliation{Department of Complexity Science and Engineering and Department of Physics, University of Tokyo, Hongo 7-3-1, Bunkyo-ku, Tokyo 113-0033, Japan}

\author{A. Fujimori}
\affiliation{Department of Complexity Science and Engineering and Department of Physics, University of Tokyo, Hongo 7-3-1, Bunkyo-ku, Tokyo 113-0033, Japan}

\author{Kouji Segawa}
\affiliation{Institute of Scientific and Industrial Research, Osaka University,
 Mihogaoka 8-1, Ibaraki, Osaka 567-0047, Japan}

\author{Yoichi Ando}
\affiliation{Institute of Scientific and Industrial Research, Osaka University,
 Mihogaoka 8-1, Ibaraki, Osaka 567-0047, Japan}

\date{\today}
\begin{abstract}

In order to study an intrinsic chemical potential jump between the hole- and electron-doped high-$T_{\rm c}$ superconductors, we have performed core-level X-ray photoemission spectroscopy (XPS) measurements of Y$_{0.38}$La$_{0.62}$Ba$_{1.74}$La$_{0.26}$Cu$_{3}$O$_{y}$ (YLBLCO), into which one can dope both holes and electrons with maintaining the same crystal structure. Unlike the case between the hole-doped system La$_{2-x}$Sr$_{x}$CuO$_{4}$ and the electron-doped system Nd$_{2-x}$Ce$_{x}$CuO$_{4}$, we have estimated the true chemical potential jump between the hole- and electron-doped YLBLCO to be $\sim$0.8 eV, which is much smaller than the optical gaps of 1.4-1.7 eV reported for the parent insulating compounds. We attribute the reduced jump to the indirect nature of the charge-excitation gap as well as to the polaronic nature of the doped carriers.

\end{abstract}
\maketitle

In condensed matter physics, chemical potential is one of the most fundamental properties and plays many important roles. For an insulator, the chemical potential is considered to show a jump between hole doping and electron doping. In a conventional intrinsic semiconductor (or band insulator), the chemical potential should be positioned at the midpoint between the conduction-band minimum (CBM) and the valence-band maximum (VBM). When a small number of holes (electrons) are doped into the intrinsic semiconductor, the chemical potential should move to the VBM (CBM) or to an impurity level formed in the vicinity of the VBM (CBM). This leads to a chemical potential jump of the magnitude nearly equal to the band gap of the undoped semiconductor. In a Mott insulator, however, the situation is far from clear. For light hole or electron doping, ``in-gap states" may be formed within the band gap in a nontrivial manner, and the chemical potential may be positioned within the gap. If this is the case, the chemical potential jump would become smaller than the band gap. In high-$T_{\rm c}$ superconductors (HTSCs), this issue has been controversial for a long time \cite{Eskes1988, Allen1990, Veenendaal1993, Veenendaal1994}. So far, although chemical potential shifts as functions of carrier dopings in HTSCs have been studied in various systems \cite{Eskes1991, Veenendaal1993, Ino1997, Harima2001, Steeneken2003, Harima2003, Yagi2006}, those studies have been made either on the hole-doped side or on the electron-doped side and have not been able to unambiguously address the question of the chemical potential jump between the hole-doped and electron-doped HTSCs. This is because, unlike some conventional semiconductors, one has not been able to dope both holes and electrons into the same parent Mott insulator. Hence, the study of the chemical potential jump in HTSCs has been performed in a rather indirect way of comparing different materials, namely, the hole-doped system La$_{2-x}$Sr$_{x}$CuO$_{4}$ (LSCO) and the electron-doped system Nd$_{2-x}$Ce$_{x}$CuO$_{4}$ (NCCO). As a result, the chemical potential jump between La$_{2}$CuO$_{4}$ (LCO) and Nd$_{2}$CuO$_{4}$ (NCO) has been found to be negligibly small or as small as $\sim$0.15-0.4 eV from the core-level X-ray photoemission spectroscopy (XPS) \cite{Harima2001} or valence-band photoemission spectroscopy studies \cite{Namatame1990, Allen1990}. Angle-resolved photoemission spectroscopy (ARPES) studies showed that the chemical potentials in LCO and NCO are located above the Zhang-Rice singlet (ZRS) band maximum by $\sim$0.4 eV \cite{Ino2000, Rosch2005} and $\sim$1.2 eV \cite{NP3}, respectively, implying a chemical potential jump in the range of 0.8 eV. According to optical studies \cite{Tokura1990, Uchida1991}, on the other hand, the band gap of LCO and NCO are found to be as large as $\sim$2.0 eV and $\sim$1.5 eV, respectively, meaning that the chemical potential jump is much smaller than the optical gaps. These results suggest that some electronic states may be formed within the optical gap upon carrier doping and pin the chemical potential. However, because LCO and NCO have different chemical compositions and different crystal structures (so-called $T$- and $T'$-structures, respectively), additional effects such as differences in Madelung potential may also complicate the comparison. Therefore, it is strongly desirable to study a $real$ chemical potential jump by investigating HTSCs that allow both hole and electron dopings in the same parent Mott insulator.

Recently, both electron and hole dopings have become possible in a new high-$T_{\rm c}$ cuprate system Y$_{0.38}$La$_{0.62}$Ba$_{1.74}$La$_{0.26}$Cu$_{3}$O$_{y}$ (YLBLCO) \cite{Segawa2006}, where some of the Y and Ba ions in YBa$_{2}$Cu$_{3}$O$_{y}$ (YBCO) are replaced by La ions, maintaining the same crystal structure as YBCO. Owing to the large in-plane lattice constant compared with that of YBCO, one can dope the system not only with holes but also with electrons through varying the oxygen content. In this Letter, we report on a core-level XPS study of hole- and electron-doped YLBLCO. The observed chemical potential jump between them is found to be $\sim$0.8 eV, much smaller than the reported optical gap of 1.4-1.7 eV of insulating YBCO \cite{Cooper1993, Takenaka1992}. We shall discuss the origin of the difference between the optical gap and the chemical potential jump.

High-quality single crystals of YLBLCO were grown by the flux method. Details can be found in Ref. \cite{Segawa2006}. An electron-doped sample ($y = 6.22$) and hole-doped samples ($y = 6.70, 6.80$, and 6.95) were prepared. Their carrier concentrations in the CuO$_{2}$ plane were 1\%, 0.3\%, 3\%, and 7\%, respectively, according to Hall-effect measurements. The changes in the carrier concentration are much smaller than those expected from the changes in $y$ probably because most of holes are removed from or enter the Cu-O chains. The XPS measurements were performed using the Mg $K\alpha$ ($h\nu=1253.6$ eV) line and a SCIENTA SES-100 electron-energy analyzer. The total energy resolution was $\sim$0.8 eV. However, owing to the highly stable power supply of the analyzer, it was possible to determine the binding energy shifts with the accuracy of $\sim$50 meV. Samples were cleaved {\it in situ} under an ultrahigh vacuum of 10$^{-10}$ Torr to obtain clean surfaces. We measured spectra at different temperatures and found that the spectra of the $y=6.22$ and $y=6.70$ samples, which had high electrical resistivities, were shifted to higher binding energies with decreasing temperature. We attributed these shifts to the charging effect and, therefore, we used only the spectra at $\sim$ 250 K for these samples. For the $y=6.80$ and $y=6.95$ samples, on the other hand, we found no charging effect down to $\sim$ 80 K, and, therefore, we used spectra taken both at $\sim$100 and at $\sim$250 K, as shown in Fig. 1(a). In fact, the observed temperature-dependent shifts for these samples were in the opposite direction to what would be expected for the charging effect and should reflect the intrinsic temperature dependence of the chemical potential in the hole-doped compounds \cite{Jaklic1996}. The 4$f$ core level of gold was used to determine the Fermi level ($E_{\rm F}$) position before and after each set of measurements.

\begin{figure}
\begin{center}
\includegraphics[width=8cm]{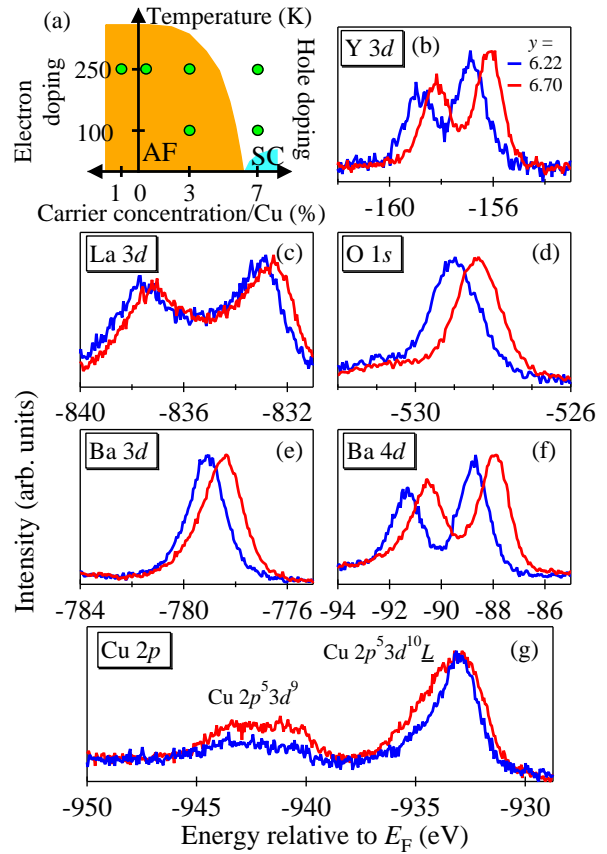}
\caption{(Color online) (a) Schematic phase diagram of Y$_{0.38}$La$_{0.62}$Ba$_{1.74}$La$_{0.26}$Cu$_{3}$O$_{y}$ (YLBLCO). Measurements were made at filled circles. Shaded regions represent the superconducting (SC) and antiferromagnetic (AF) phases. (b)-(g) XPS spectra of electron-doped $y = 6.22$ and hole-doped $y = 6.70$ YLBLCO at 250 K for the Y 3$d$, La 3$d$, O 1$s$, Ba 3$d$, Ba 4$d$, and Cu 2$p$ core levels. For the Cu 2$p$ XPS spectra, the peak around $-933.5$ eV is the main peak due to 2$p^{5}$3$d^{10}\underline{L}$ final states, where $\underline{L}$ represents the ligand hole, and the broad peak around $-943$ eV is the charge-transfer satellite due to 2$p^{5}$3$d^{9}$ final states.}
\end{center}
\end{figure}

Figure 1(b)-(g) shows the core-level XPS spectra of the electron-doped $y = 6.22$ and hole-doped $y = 6.70$ YLBLCO samples taken at 250 K. In Fig. 1(g), the main peak of the Cu 2$p_{3/2}$ spectrum (corresponding to 2$p^{5}$3$d^{10}\underline{L}$ final states, where $\underline{L}$ represents a ligand hole) is located at $-933.5$ eV close to those of LCO and NCO in the previous studies \cite{Cummins1993, Koitzsch2002, Taguchi2005}. The main peak width for the $y = 6.22$ sample is narrower than that for the $y = 6.70$ sample. Since the main peak of Cu$^{+}$ is known to be sharper than that of Cu$^{2+}$, the variation of the peak width is attributed to the replacement of Cu$^{2+}$ by Cu$^{+}$. Also, the intensity of the charge-transfer satellite around $-943$ eV (2$p^{5}$3$d^{9}$ final state) for $y = 6.22$ is weaker than that for $y = 6.70$. Since the satellite peak has been observed in the Cu 2$p$ XPS spectra of Cu$^{2+}$ compounds but not in those of Cu$^{+}$ compounds \cite{Ghijsen1988}, the suppression also suggests the replacement of Cu$^{2+}$ by Cu$^{+}$. These results indicate that electrons are indeed doped into the system. The large change in the Cu valence for a small change in the carrier concentration of the CuO$_{2}$ plane indicates that the holes doped by oxygen are localized in the Cu-O chain. The XPS spectra of the Y 3$d$, La 3$d$, O 1$s$, Ba 3$d$, and Ba 4$d$ core levels of the $y = 6.22$ sample are shifted to higher binding energies than those of the $y = 6.70$ sample nearly by the same amount. As for the Ba 3$d$, Ba 4$d$, and Cu 2$p$ core levels, their line shapes change with $y$, indicating that it is difficult to estimate the chemical potential shift using these core-level spectra. Therefore, the Y 3$d$, La 3$d$, and O 1$s$ core levels are more suitable for estimating the chemical potential shift.

\begin{figure}
\begin{center}
\includegraphics[width=8.7cm]{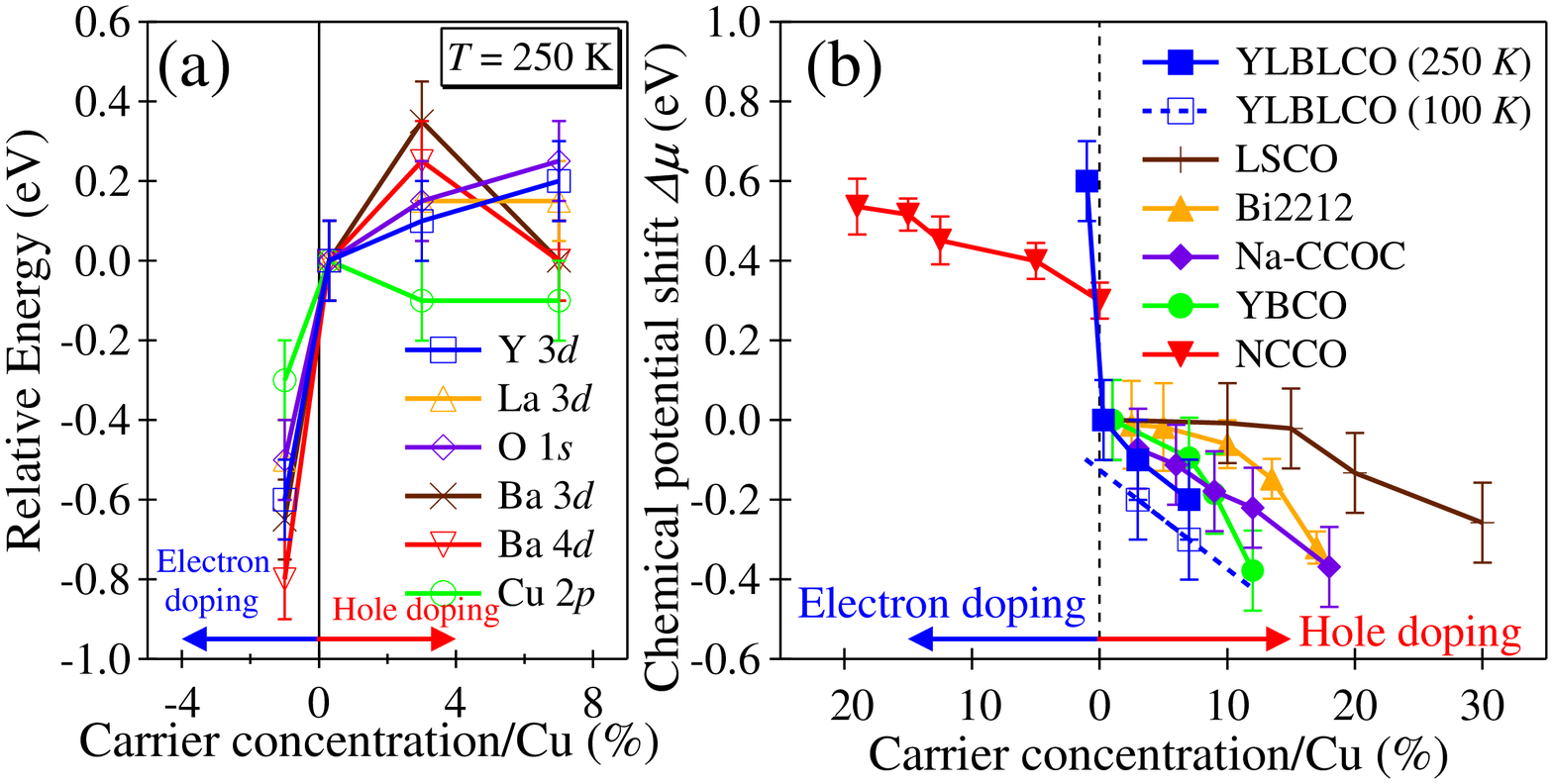}
\caption{(Color online) (a) Energy shift of each core level in YLBLCO relative to $y = 6.70$ at 250 K as a function of carrier concentration. (b) Chemical potential shifts in YLBLCO at 250 and 100 K, compared with those in La$_{2-x}$Sr$_{x}$CuO$_{4}$ (LSCO) \cite{Ino1997}, Bi$_{2}$Sr$_{2}$Ca$_{1-x}$(Pr, Eu)$_{x}$Cu$_{2}$O$_{8+ \delta}$ (Bi2212) \cite{Harima2003}, Na$_{x}$Ca$_{2-x}$CuO$_{2}$Cl$_{2}$ (Na-CCOC) \cite{Yagi2006}, YBCO \cite{Yagi}, and Nd$_{2-x}$Ce$_{x}$CuO$_{4}$ (NCCO) \cite{Harima2001}. Note that the shift for NCCO relative to the hole-doped systems are rather arbitrary \cite{Harima2001}.}
\end{center}
\end{figure}

Figure 2(a) summarizes the energy shift $\Delta E$ of each core level relative to that of the $y = 6.70$ YLBLCO sample. We have estimated the core-level shifts using both the peak position and the low binding energy side of the peak for each core level to check internal consistency. While the Y 3$d$, La 3$d$, and O 1$s$ core levels show similar shifts, the Cu 2$p$, Ba 3$d$, and Ba 4$d$ core levels show different shifts, due to the changes in the spectra-line shapes as shown in Fig. 1. When the band filling is varied, the change in the core-level energy relative to $E_{\rm F}$, $\Delta E$, is generally given by $ \Delta E = -\Delta \mu -K \Delta Q-\Delta V_{M}+\Delta E_{R}$ \cite{Ino1997, Harima2001}, where $\Delta \mu$ is the change in the chemical potential, $\Delta Q$ is the change in the number of valence electrons on the atom and $K$ is a constant, with $-K\Delta Q$ presenting the so-called chemical shift, $\Delta V_{M}$ is the change in the Madelung potential, and $\Delta E_{R}$ is the change in the extra-atomic relaxation energy. When carriers are doped into the CuO$_{2}$ plane, $K\Delta Q$ changes in proportion to the carrier concentration. We observed such a behavior only for Cu 2$p_{3/2}$ and could ignore $K\Delta Q$ for Y 3$d$, La 3$d$, and O 1$s$ core levels. $\Delta V_{M}$ is negligibly small as in other transition-metal oxides \cite{Fujimori2002} because the core levels of cations (Y 3$d$ and La 3$d$) and an anion (O 1$s$) move in the same direction by the same amount. $\Delta E_{R}$ is due to the change in the screening of the core hole potential and, therefore, should be larger for atoms in the metallic CuO$_{2}$ plane than for those in the insulating block layers. The same shifts of the Y 3$d$, La 3$d$, and O 1$s$ core levels suggest that the $\Delta E_{R}$ term is also negligibly small. Indeed, in the previous reports on the chemical potential shift in filling control oxides \cite{Ino1997, Harima2001, Yagi2006}, the effect of $\Delta V_{M}$ and $\Delta E_{R}$ could be ignored. Therefore, we consider that the core-level shifts of Y 3$d$, La 3$d$, and O 1$s$ reflect the chemical potential shift $\Delta \mu$, and $\Delta \mu$ for YLBLCO has been estimated from the average of the Y 3$d$, La 3$d$, and O 1$s$ core-level shifts.

\begin{figure}
\begin{center}
\includegraphics[width=8.5cm]{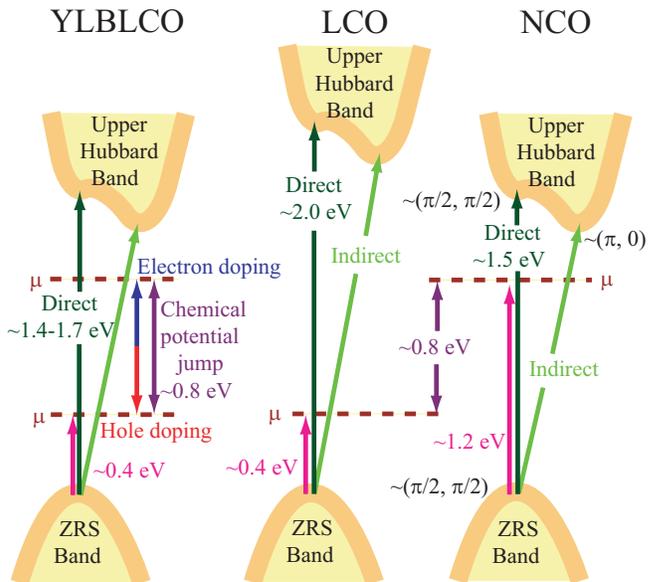}
\caption{(Color online) Schematic band diagram of YLBLCO and those of LCO and NCO. Orange (thick gray) curves represent the schematic dispersions of energy distribution curve peaks. The positions of the chemical potential $\mu$ are indicated by horizontal dashed lines. The maximum positions of the Zhang-Rice singlet (ZRS) band are aligned. For LCO and NCO, the energy separation between the ZRS band and the chemical potential was obtained from angle-resolved photoemission studies \cite{Ino2000, Rosch2005, NP3}. The magnitude of the direct gap was estimated from optical studies \cite{Tokura1990, Uchida1991}. For YLBLCO, the maximum position of the ZRS band relative to the chemical potential is assumed to be the same as those of lightly-doped LSCO and Na-CCOC \cite{Ino2000, Rosch2005, Shen2004}. The direct gap of YLBLCO is estimated from the optical studies of YBCO \cite{Cooper1993} and PrBa$_{2}$Cu$_{3}$O$_{y}$ \cite{Takenaka1992}.}
\end{center}
\end{figure}

Figure 2(b) shows the chemical potential shift of YLBLCO at 250 K and 100 K thus deduced as well as those of La$_{2-x}$Sr$_{x}$CuO$_{4}$ (LSCO) \cite{Ino1997}, Nd$_{2-x}$Ce$_{x}$CuO$_{4}$ (NCCO) \cite{Harima2001}, Bi$_{2}$Sr$_{2}$Ca$_{1-x}$(Pr, Eu)$_{x}$Cu$_{2}$O$_{8+ \delta}$ (Bi2212) \cite{Harima2003}, Na$_{x}$Ca$_{2-x}$CuO$_{2}$Cl$_{2}$ (Na-CCOC) \cite{Yagi2006}, and YBCO \cite{Yagi}. On the hole-doped side, the behavior of YLBLCO is similar to those of Bi2212, Na-CCOC, and YBCO, but is different from that of LSCO. This difference has been attributed to the small next-nearest-neighbor hopping $t'$ or stripe effects in LSCO as discussed previously \cite{Tohyama2003, Ino1997, Yamada1998}, indicating that YLBLCO has no effect of small $t'$ or stripes. Between the hole-doped and electron-doped sides of YLBLCO, a chemical potential jump of $\sim$0.6 eV is observed at 250 K. However, since a chemical potential shift of $\sim$0.1 eV was observed between 250 K and 100 K in some hole-doped samples, the chemical potential jump would be larger at low temperatures. If the temperature dependent shift is symmetric between the electron-doped and hole-doped sides, the chemical potential jump in YLBLCO at 100 K would be estimated to be larger than that at 250 K by $\sim$0.2 eV, that is, $\sim$0.8 eV. This value is small compared with the gap ($\sim$1.4-1.7 eV) estimated from the optical studies \cite{Cooper1993, Takenaka1992}, but is comparable to the activation gap ($\sim0.9$ eV) determined from high-temperature transport \cite{Ono2007}. The chemical potential jump of YLBLCO is somewhat larger than that estimated for LCO and NCO from the core-level XPS study \cite{Harima2001} and from the energy difference of the band structure ($\sim$0.3 eV) \cite{Ikeda20092}. On the other hand, the energy difference between the ZRS band maximum and the chemical potential for LCO and NCO was found to be $\sim$0.4 eV \cite{Ino2000, Rosch2005} and $\sim$1.2 eV \cite{NP3}, respectively. Therefore, if the energy positions of the ZRS band are taken as the energy reference, the chemical potential jump would be $\sim$0.8 eV. Generally, the magnitude of the optical gap in the charge-transfer-type compounds is influenced by the Madelung potential. The fact that the optical gap in LCO is larger than that in NCO by $\sim$0.5 eV \cite{Uchida1991, Tokura1990} is caused by the Madelung-potential difference due to the different crystal structures, meaning that $\Delta V_{M}$ is not negligible in the estimation of $\Delta \mu$ from $\Delta E$. Thus, we conclude that the intrinsic chemical potential jump between hole- and electron-doped HTSCs is $\sim$0.8 eV, and the different Madelung potential between LCO and NCO makes it difficult to estimate the chemical potential jump between them.

Finally, we summarize the energy diagram of YLBLCO, LCO, and NCO in Fig. 3. The energy difference of $\sim$0.4 eV between the VBM and the chemical potential for LCO has been attributed to a polaronic effect \cite{Rosch2004, Mishchenko2004}, which shifts the (multi-phonon) peak in energy distribution curves toward higher binding energies through the Frank-Condon principle. We should also note that the minimum charge-transfer gap is not a direct gap which was detected in the optical studies but an indirect one as demonstrated by resonant inelastic X-ray scattering experiments \cite{Hasan2000} combined with extended Hubbard-model calculation \cite{Tsutsui1999}, according to which the indirect gap from $\vec{k}\sim(\pi/2, \pi/2)$ to $\sim(\pi, 0)$ is found to be smaller than the direct one $\sim(\pi/2, \pi/2)$.

In conclusion, we have performed core-level XPS measurements of YLBLCO in order to study the chemical potential jump between the hole-doped and electron-doped sides. Unlike the case of LCO and NCO, the YLBLCO results yield the true chemical potential jump between the hole- and electron-doped HTSCs. The deduced chemical potential jump ($\sim$0.8 eV) is much smaller than that of the optical gap ($\sim$1.4-1.7 eV), but is comparable to the activation gap ($\sim$0.9 eV) determined from the high-temperature transport. We attribute the smallness of the jump to the indirect nature of the charge-excitation gap as well as to a polaronic effect on the doped carriers.

We are grateful to H. Yagi for discussion. This work was supported by KAKENHI 19204037. M. I. was supported by the Global COE Program, MEXT, Japan, Y. A. by KAKENHI 19674002 and 20030004, and K. S. by KAKENHI 20740196.

\end{document}